\documentstyle[12pt]{article}
\begin{document}

\title
{\Large \bf On  Synchronization With Positive Conditional Lyapunov Exponents}

\author{ Changsong Zhou$^1$ and  C.-H. Lai$^{1,2}$ \\
        $^1$Department of Computational Science\\
        and $^2$Department of Physics\\
        National University of Singapore,
        Singapore 119260}

\date{}
\maketitle

\begin{center}
\begin{minipage}{14cm}

\centerline{\bf Abstract}

Synchronization of chaotic system may occur only when the largest
conditional Lyapunov exponent of the driven system is negative.  The
synchronization with positive conditional Lyapunov reported in a recent
paper (Phys. Rev. E, {\bf 56}, 2272 (1997)) is a combined result of
the contracting region of the system and the finite precision in computer
simulations.

\bigskip
\baselineskip 24pt

PACS number(s): 05.45.+b;

\end{minipage}
\end{center}

\newpage

Sensitivity to initial conditions is a generic feature of chaotic
dynamical systems. Two chaotic orbits, starting from slightly different
initial points in the state space, separate exponentially with time,
and become totally uncorrelated.  As a result, independent identical
chaotic systems cannot synchronize with each other.  The sensitivity is
quantitatively described by positive  Lyapunov exponent(s) in the
Lyapunov exponent spectrum of the chaotic system.

However, chaotic systems linked by common signal can synchronize with
each other. Several cases could  be distinguished.  In the first case,
a replica  subsystem driven by  chaotic signals of the
chaotic system can synchronize identically with the drive system[1-5],
if the largest conditional Lyapunov is negative. This is
referred to as {\sl identical synchronization}.

Secondly, a driven system, which is not a replica of the drive system,
however, may not achieve identical synchronization, but {\sl
generalized synchronization}[6-8], if the largest conditional Lyapunov 
exponent is negative. 
Two identical systems, driven by the same signal, thus may
come to the same final state due to the  negative largest conditional
Lyapunov exponent.

Lyapunov exponents are also employed to characterize behavior 
of  random dynamical systems[9]: the system is chaotic (non-chaotic)
when the largest Lyapunov exponent is positive (negative).
The sensitivity of a  chaotic system may also be 
suppressed by noise, and  identical chaotic systems subjected to common
noise can synchronize with each other.
Maritan and Banavar[10] studied the behavior of the noise-driven 
logistic maps and reported synchronization phenomenon.
It turned out that the observed synchronization was an outcome of
finite precision in numerical simulations[11,12], while the Lyapunov
exponent of the noisy logistic map is positive[11].

Very recently,  Shuai {\sl et al}[13] claimed that
synchronization can be achieved with positive conditional Lyapunov
exponents. In a one-way coupled map lattice, they observed, through
computer simulations,  synchronization of spatiotemporal chaos  with
many positive components in the conditional  Lyapunov exponent
spectrum.  Based on these results, they drew the conclusion that the
conditional Lyapunov exponents cannot be used as a criterion for
synchronous chaotic systems.

Whether such a claim is true is of great importance for our
understanding of synchronization.  In this paper, we reexamined  such
synchronization phenomenon, revealing that it is yet another example of
round-off induced phenomenon.

In[13], Shuai {\sl et al\/} studied a driven one-way coupled  map lattice
\begin{eqnarray}
y_i(t+1)=(1-\epsilon)f(y_i(t))+\epsilon f(y_{i+1}(t)) \;\;\;\;
(i=1,\cdots, N),\\ 
y_{N+1}(t)=x_{0}(t),           
\end{eqnarray}
where $x_0(t)$ is a hyperchaotic signal from a one-way coupled ring lattice
\begin{eqnarray}
x_{0}(t+1)=(1-\epsilon_0)f(x_0(t))+\epsilon_0 f(x_1(t)),\\
x_i(t+1)=(1-\epsilon)f(x_i(t))+\epsilon f(x_{i+1}(t))  \;\;\;\;
(i=1,\cdots, N),\\ 
x_{N+1}(t)=x_{0}(t),
\end{eqnarray}
The chaotic map is the well-known logistic map $f(x)=4x(1-x)$ and
$\epsilon_0=0.01$ 

As in [13],  the conditional Lyapunov exponents of the driven system are
\begin{equation}
\lambda_i=\ln(1-\epsilon)+\lim_{T\to
\infty}\frac{1}{T}\sum\limits_{t=1}^{T}|f^{\prime}(y_i(t))| 
\end{equation}

Let us study the simplest case of $N=1$. The conditional Lyapunov exponent 
as a
function of $\epsilon$ is shown in Fig. 1. To detect the behavior of
synchronization, 100 performances with random initial conditions are
carried out for each $\epsilon$.  Synchronization occurs when $y_1$ and
$x_1$ become numerically identical for the finite precision in
simulations (double precision). $P=M/100$, where $M$ is the number of
simulations in which synchronization occurs within $5\times 10^7$
iterations, is estimated as a function of $\epsilon$, as shown in
Fig.~1.  It can be detected that synchronization with positive Lyapunov
exponent (SP) occurs in several regions. We will take $\epsilon=0.200$
and $\epsilon=0.335$ as examples, as was pointed out in [13].

Is SP a true physical phenomenon or an artifact of finite precision in
computer simulations?  In the following, different precision formats (single,
double and quadruple precision) are employed in the simulations. Firstly,
the difference $e(t)=|y_1(t)-x_1(t)|$ preceding the synchronized state
is examined for simulations with different precisions but the same
random initial conditions. The results for $\epsilon=0.200
(\lambda_1=0.105)$ and $\epsilon=0.335 (\lambda_1=0.025)$ are  shown in
Fig. 2(a) and (b),  respectively.  For comparison, an example of
synchronization with negative conditional Lyapunov exponent (SN)  at
$\epsilon=0.520 (\lambda_1=-0.041)$ is illustrated in Fig. 2(c). Note
the different scales for the different precisions used.
$e(t)$ displays  an intermittent behavior before reaching SP.  SP
occurs somewhat abruptly, when $e(t)$  drops lower than the precision
of the computer. The time $T$ needed for SP to occur is  much longer
for quadrupole precision  than that for single and double precisions.
As for SN, $e(t)$  continues its trend of decrease when higher
precision is employed in the simulation. It is plausible to imagine
that for SP, the intermittence of $e(t)$  will continue indefinitely for
infinite precision, while for SN, $e(t)$  will approach  to $0$. So, the
physical process of SP is an intermittence, with $e(t)$  becoming very
small and enlarging to the size of the chaotic attractor alternately.

The difference $e(t)$ is actually not zero even beyond the precision of
the computer. In the following simulation with quadrupole precision,
when the states of the systems are numerically identical (SP), a
perturbation $\xi \in (-10^{-30}, 10^{-30})$ is added  to the drive
signal $x_0(t)$ of system $y$ at the next iteration, under the
constrain  $0<x_0(t)+\xi<1$. Such a tiny perturbation can totally
destroy the synchronization behavior when $\lambda_1>0$, as seen from
Fig. 3(a) and (b) for the results of $\epsilon=0.200$ and
$\epsilon=0.335$ respectively, because the tiny difference can be
amplified to the order of $10^0$ due to the positive conditional
Lyapunov. While for SN, $e(t)$  continues to decrease after the
impulsive perturbations, and the level of difference is the order of
$10^{-30}$. Such a dramatic difference between the behavior of SP and
SN shows that, negative conditional Lyapunov is necessary condition for
physical synchronization.
 
To demonstrate further that synchronization cannot be achieved
physically with a positive conditional Lyapunov exponent, the average
synchronization time $T_a$ is evaluated for different precisions with
100 random initial conditions. The results for SP ($\epsilon=0.200$ and
$\epsilon=0.335$) are displaced with a linear-log plot in Fig. 4(a).
The three points lies almost on a straight line, meaning that $T_a\sim
\exp(AL)$, where $L$  is  the number of significant digits of the
finite precision.  An exponential increase of $T_a$ with $L$ proves that
synchronization can never occur with infinite precision. The behavior
of SN is greatly different, where $T_a$ follows a linear dependence on $L$,
$T_a\sim BL$, as seen from the result of $\epsilon=0.52$ displaced with
a linear-linear plot in Fig. 4(b).  The  reason is  that,
approximately,  $e(t)$  decreases exponentially with time, so that
$10^{-L}\sim \exp(\lambda_1T_a)$, resulting in $B=-\ln 10/\lambda_1.$
$B=56.2$ at $\epsilon=0.520$ is in good agreement with the slope 58.5
of the solid line in Fig. 4(b).
 
Why is then that SP can be observed in numerical simulations even
within thousands of iterations?  The origin is that there are
contracting regions in a map $f$, $C=\{x, y | |f(x)-f(y)|<1\}$.  Two
orbits in a contracting region come closer to each other at the next
step. For the  system studied above, the contracting region is
$1-\frac{1}{4(1-\epsilon)}<x_1+y_1<1+\frac{1}{4(1-\epsilon)}$.  The
strip near $x_1+y_1=1$ has the strongest contracting rate.  The
distribution of $x_1+y_1$ is calculated with $10^7$ iterations to
examine the relationship between SP and $C$. As seen from the results
of $\epsilon=0.200$ and $\epsilon=0.335$ in Fig. 5(a), the distribution
for $\epsilon=0.200$ has  very high peaks in the contracting region,
while a lower peak for $\epsilon=0.335$. Such greater frequency of access
to the contracting region at $\epsilon=0.200$ makes  $e(t)$ drops to
a much small value more frequently than at $\epsilon=0.335$
(see Fig. 3), which accounts for the result that SP is observed at
$\epsilon=0.2$ with much fewer iterations than that at
$\epsilon=0.335$. However, the evolution of the difference $e(t)$  is a
combined result of local stability and instability.  The finite-time
Lyapunov exponent [14]
\begin{equation}
\lambda^{(m)}=\frac{1}{m}\sum\limits_{t=1}^m \ln|f^{\prime}(y_1(t))|
\end{equation}
measures the average expansion or contraction rate in $m$ steps.  The
distributions of $\lambda^{(m)} (m=70)$ for $\epsilon=0.200$ and
$\epsilon=0.335$ are illustrated in Fig. 5(b). A pronounced tail to
$-1.0$ at $\epsilon=0.200$ means that the difference shrinks by a
factor of $e^{-70}\doteq 4\times 10^{-31}$ in some successive 70
iterations. The negative tails thus plays an important role in
observation of SP. It also explains the fact that SP is easier to occur
at $\epsilon=0.200$ with a larger positive conditional Lyapunov
exponent $(\lambda=0.105)$ than at $\epsilon=0.335$ ($\lambda=0.025$).
The distributions of $\lambda^{(m)}$ also reflects the true dynamics
of SP: the difference $e(t)$ can be  very small in a period of time,
and it will be amplified in some other period of time because temporal
separation dominates, thus resulting  in an
intermittent dynamics.  So the finite-time Lyapunov exponent gives a
more convincing account for the occurrence of SP in simulations.

Based on the above analysis of the simplest case of $N=1$, the SP
observed in computer simulations is a combination of two factors: the
shift of the state of the chaotic system to the  contracting region and
the finite precision in numerical simulations. For the case of $N>1$,
SP is observed for similar reasons.  The driven system is coupled in a
cascade way $x_N\to x_{N-1}\to \cdots \to x_1$, and  synchronization
can only occur for the first several $N_1$ nodes if all the $N_1$
conditional Lyapunov exponents $\lambda_N, \lambda_{N-1},\cdots
\lambda_{N-N_1+1}$ are negative. Physical synchronization of all the
lattices can only occur when all the conditional Lyapunov exponents are
negative.

In conclusion, synchronization with positive conditional Lyapunov
exponents in computer simulations is a round-off induced phenomenon.
The physical dynamics of SP is an intermittence.  One can expect to
observe SP easily in computer simulations  in such systems with
large contracting regions, and the couplings have the effect of
shifting the state to the contracting regions so that the finite-time
Lyapunov has a significant tail of negative values.  A negative 
 conditional
Lyapunov exponent is a  necessary condition for synchronizing  chaotic
systems.

\bigskip
{\bf Acknowledgements:}
This work was supported in part by research grant RP960689 at the National
University of Singapore.  CZ is a NSTB Postdoctoral Research Fellow.

\newpage
\bigskip

{\bf References}
\begin{description}
\item [1.] L.M. Pecora and T.L. Carroll,
          Phys. Rev. Lett. {\bf 64}, 821 (1990);
          Phys. Rev. {\bf 44A}, 2374 (1991).
\item [2.] Kocarev and U. Parlitz, Phys. Rev. Lett. {\bf 74}, 5028 (1995).
\item [3.] K. Pyragas, Phys. lett. {\bf 181A}, 203 (1993).
\item [4.] M. Ding and E. Ott, Phys. Rev. E {\bf 49}, R945 (1994);
\item [5.] T. L. Carroll, J. F. Heagy, and L. M. Pecora, Phys. Rev. E {\bf 54}, 4676 (1996).
\item [6.] N. F. Rulkov, M. M. Sushchik and L. S. Tsimring, Phys. Rev. E {\bf 51}, 980 (1995)   
\item [7.] H. D. I. Abarbanel, N. F. Rolkov, and M. M Sushchik, Phys. Rev. E {\bf 53}, 4528 (1996).
\item [8.] L. Kocarev and U. Parlitz, Phys. Rev. Lett. {\bf 76}, 1816 (1996).
\item [9.] Lei Yu, E. Ott, Qi Chen, Phys. Rev. Letts. {\bf 65}, 2935 (1990).
\item [10.]  A. Maritan and J R. Banavar, Phys. Rev. Letts. {\bf 72}, 1451 (1994).
\item [11.]  A. S. Pikovsky, Phys. Rev. Letts. {\bf 73} 2931 (1994).
\item [12.]  L. Longa, E. M. F. Curado, and F. A. Oliveira, Phys. Rev. E {\bf 54}, R2201 (1996)
\item [13.] J. W. Shuai, K. W. Wong, and L. M. Cheng, Phys. Rev. E. {\bf 56}, 2272 (1997). 
\item [14.] H. Herzel and B. Pompe, Phys. Lett A {\bf 122}, 121 (1983).
\end{description}

\newpage
\bigskip

{\large \bf Figure Captions}
\begin{description}
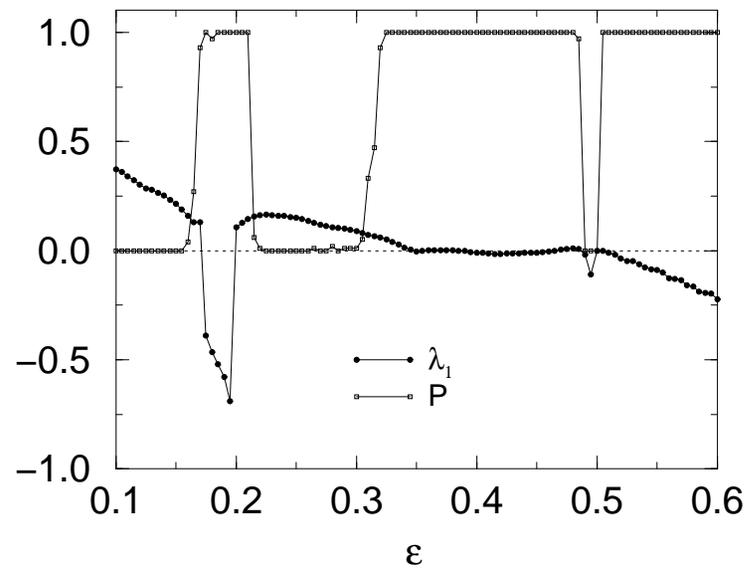
\item Fig. 1. Conditional Lyapunov exponent $\lambda_1$ and
synchronization ratio $P$ as functions of $\epsilon$.

\item Fig. 2. The time series of the difference $e(t)$ proceeding
synchronization in simulations with single, double and quadrupole
precisions. (a)$\epsilon=0.200$, SP; (b) $\epsilon=0.335$, SP; and
(c)$\epsilon=0.52$, SN.

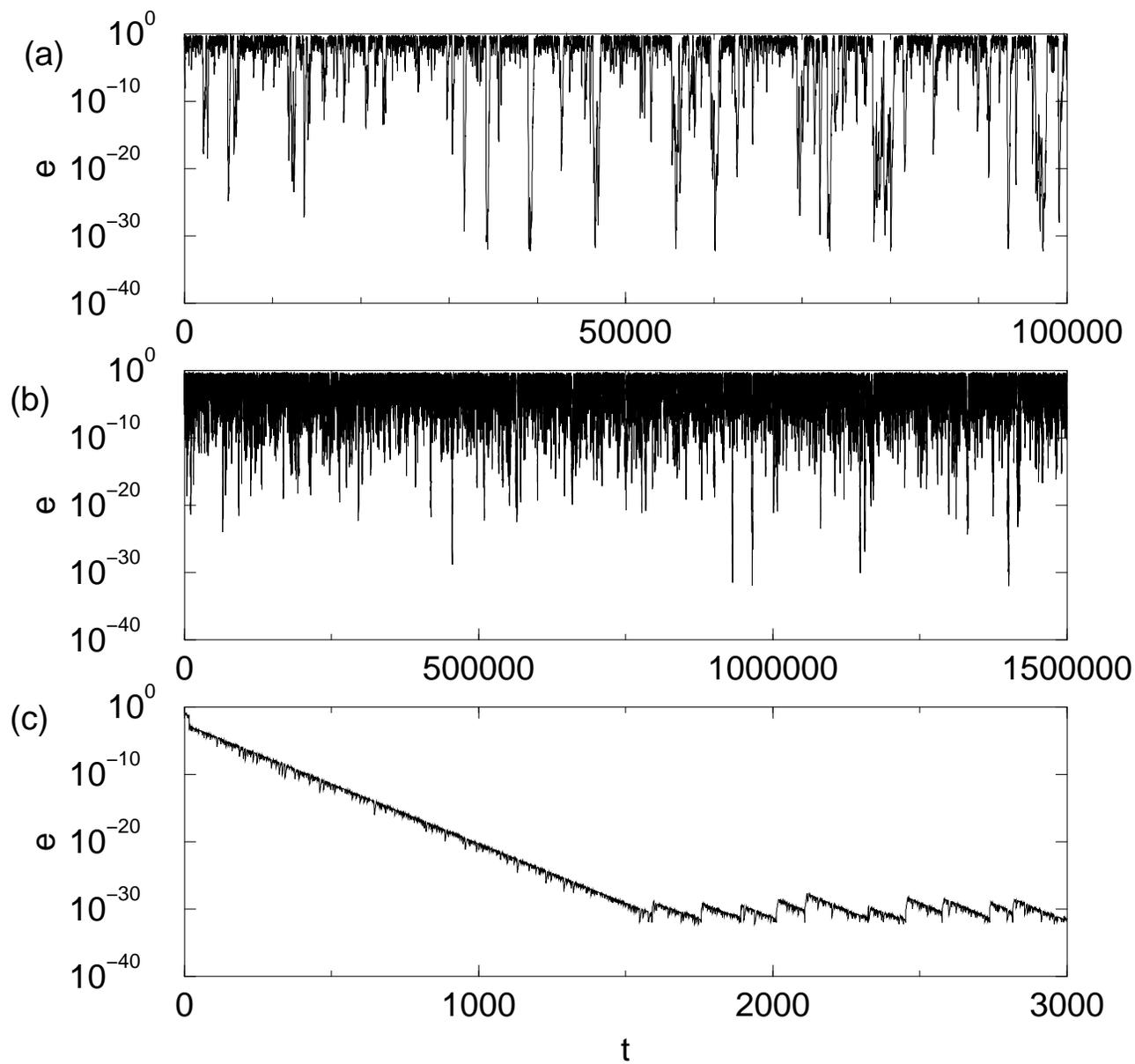
\item Fig. 3. The time series of difference $e(t)$ under impulsive
perturbations between $(10^{-30}, 10^{-30})$ in  simulations with
quadrupole precision. The initial conditions are the same as Fig. 2.
(a)$\epsilon=0.200$, SP; (b) $\epsilon=0.335$, SP; and
(c)$\epsilon=0.52$, SN.

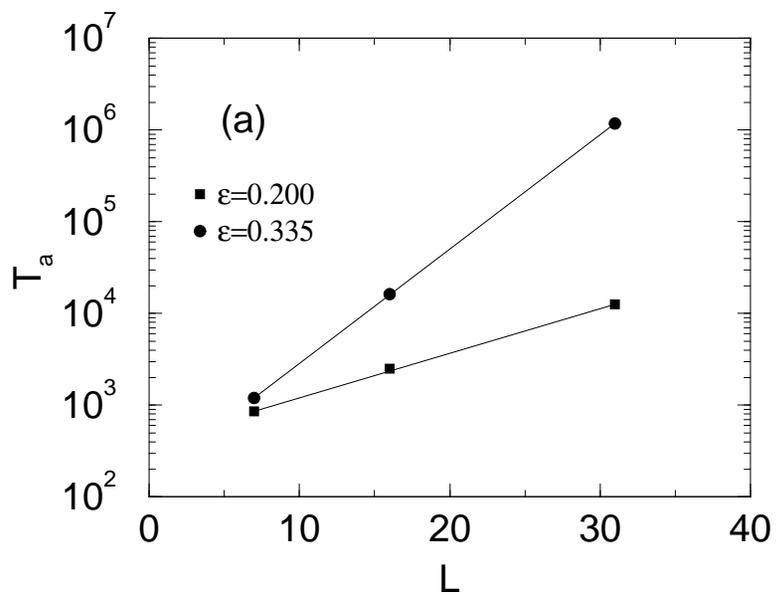
\item Fig. 4. Average synchronization time $T_a$ as a function of
precision in simulations.  $L=7, 16$ and 31 for single, double and
quadrupole precision, respectively. The each solid line links  the
first and last point of each data set. (a) Linear-log plots for SP.
(b)A linear-linear plot for SN.

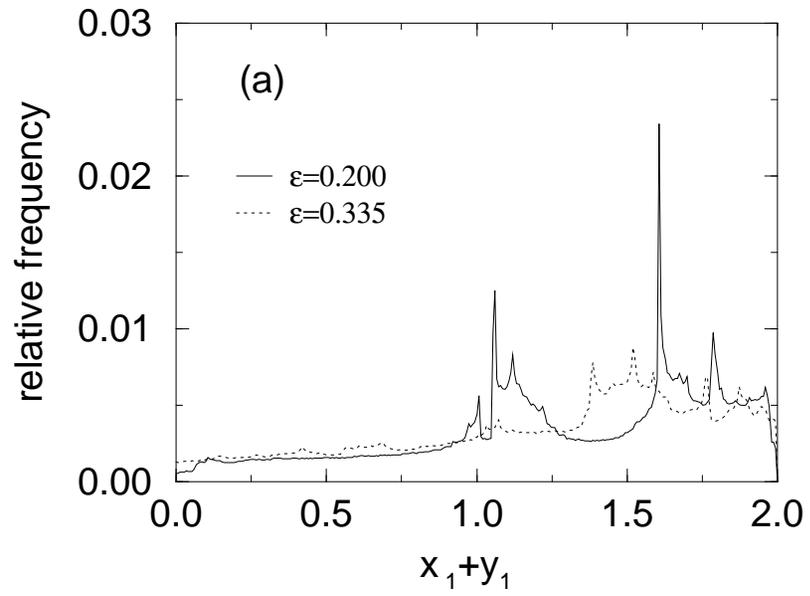
\item Fig. 5. (a) Normalized histograms of $x_1+y_1$. (b) Normalized
histograms of finite-time Lyapunov exponent.

\end{description}  
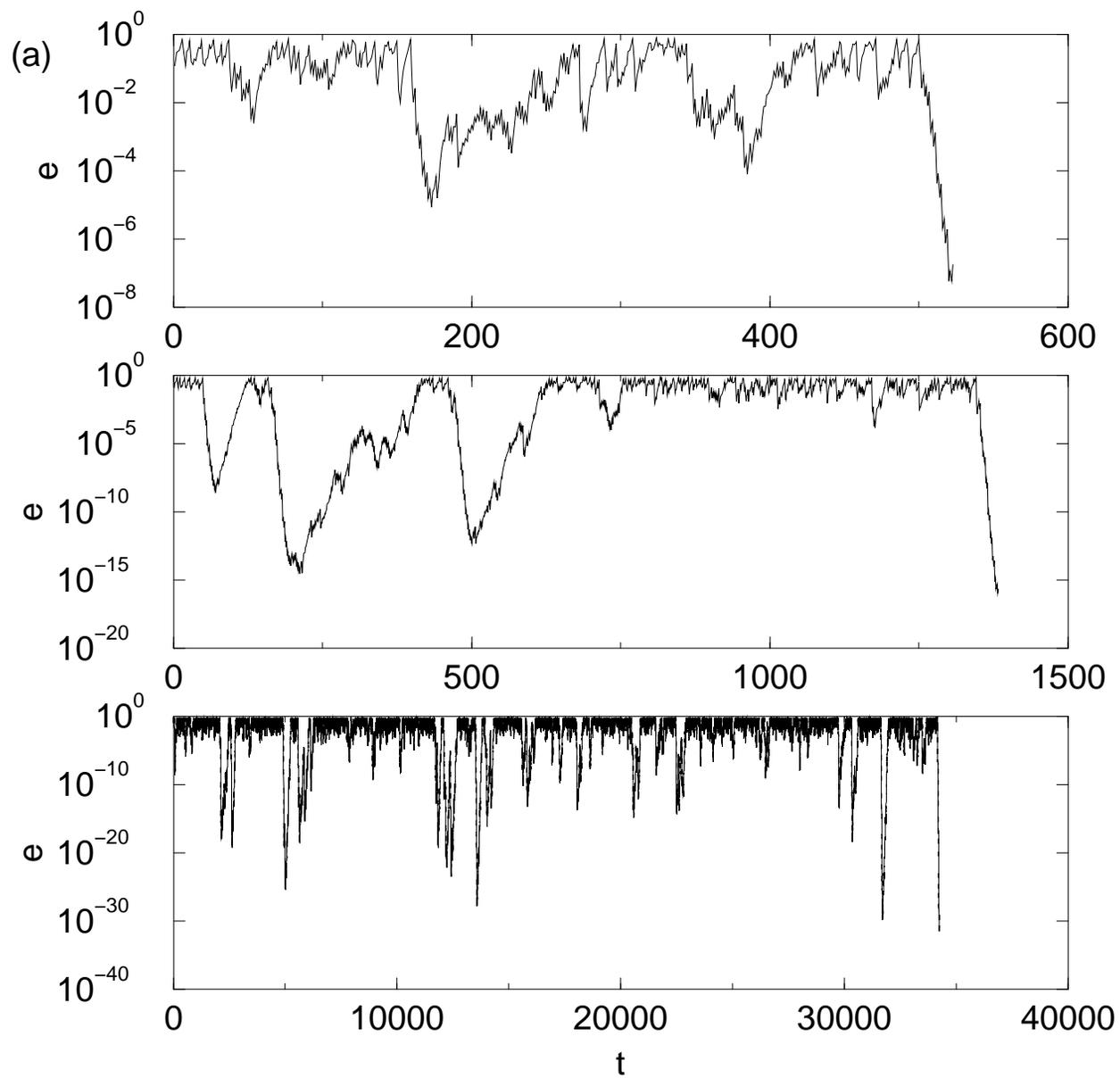
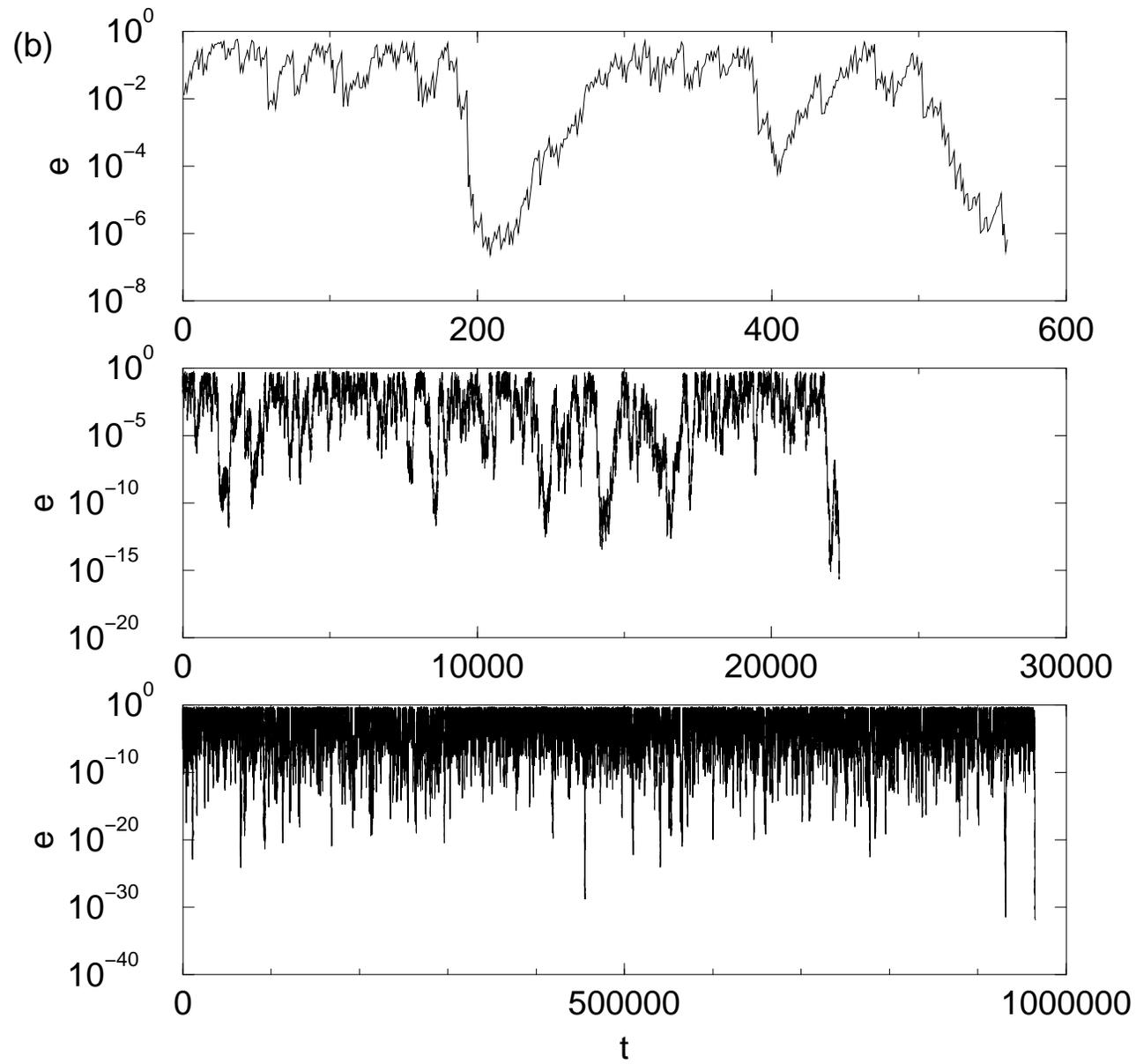
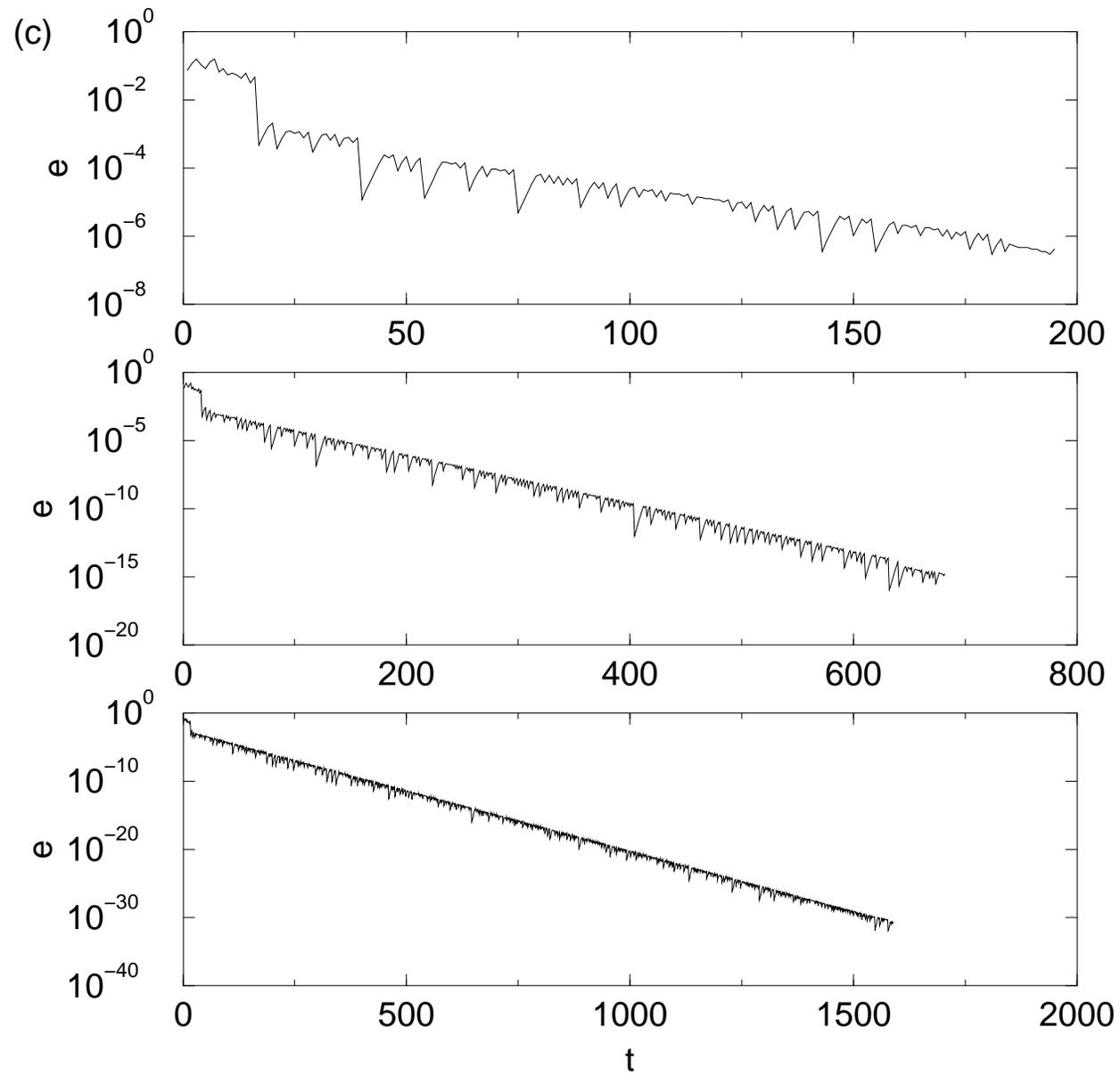
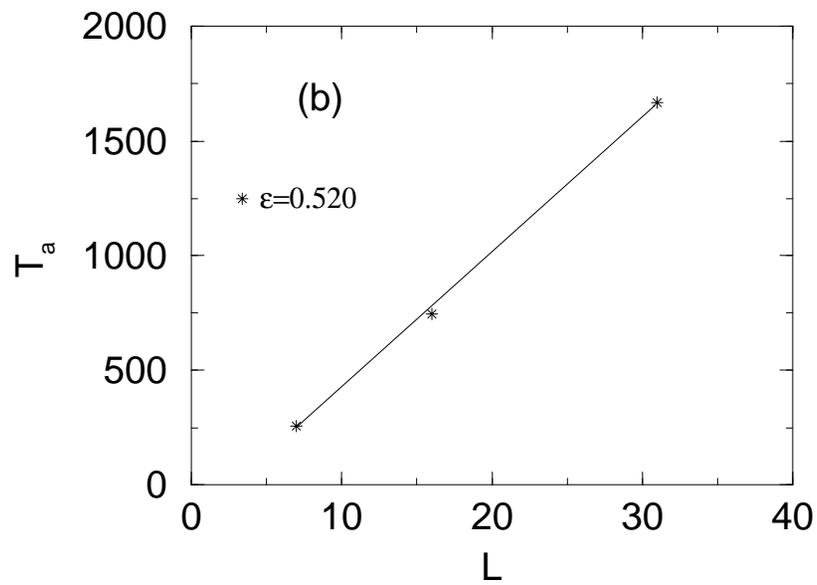
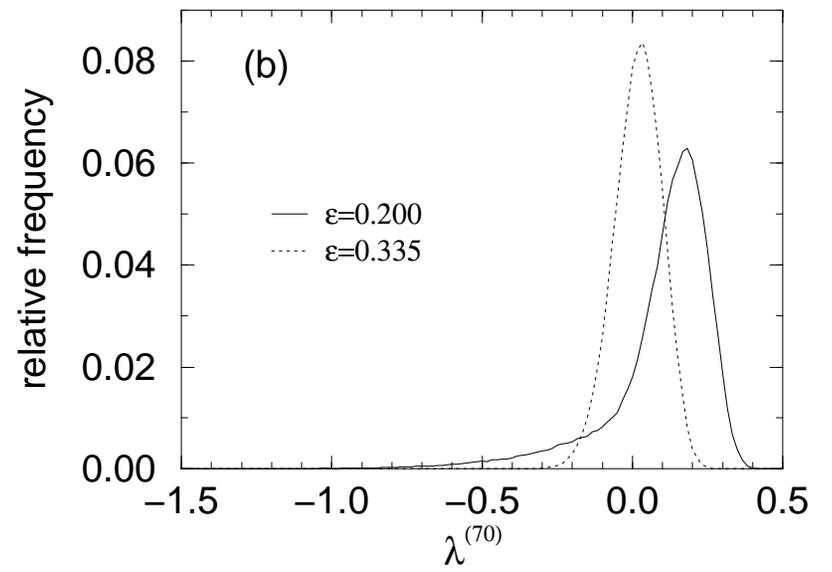
\end{document}